\documentclass[aps,prl,twocolumn,showpacs,superscriptaddress]{revtex4}
\usepackage{amsmath,amssymb,graphicx}

\begin{document}


\title{Shape diagram of vesicles in Poiseuille flow}


\author{Gwennou Coupier}
\email[]{gwennou.coupier@ujf-grenoble.fr}

\author{Alexander Farutin}

\affiliation{Laboratoire Interdisciplinaire de Physique, CNRS et Universit\'e~J.~Fourier~-~Grenoble~I, BP 87, 38402 Saint-Martin d'H\`eres, France}

\author{Christophe Minetti}

\affiliation{Microgravity Research Center, Universit\'e Libre de Bruxelles, 50 Av. F. Roosevelt, B-1050 Brussels, Belgium}

\author{Thomas Podgorski}

\author{Chaouqi Misbah}

\affiliation{Laboratoire Interdisciplinaire de Physique, CNRS et Universit\'e~J.~Fourier~-~Grenoble~I, BP 87, 38402 Saint-Martin d'H\`eres, France}
\date{\today}

\begin{abstract}

Soft bodies flowing in a channel often exhibit parachute-like shapes usually attributed to an increase of hydrodynamic constraint (viscous stress and/or confinement). We show that the presence of a fluid membrane leads to the reverse phenomenon and build a phase diagram of shapes --- which  are classified as bullet, croissant and parachute --- in channels of varying aspect ratio. Unexpectedly, shapes  are relatively wider in the narrowest direction of the channel. We highlight the role of flow patterns on the membrane in this response to the asymmetry of stress distribution.

 \end{abstract}


\maketitle

The shape of soft bodies under flow is governed by strong non-linear coupling between hydrodynamic stresses and elastic restoring forces. The latter are often linked with specific interface properties, like surface tension, or bending and shear elasticity of an elastic or liquid membrane.

A widely studied flow is the confined Poiseuille flow, in which the behavior of  red blood cells (RBCs) \cite{skalak69,gaehtgens80,mcwhirter09,mcwhirter11}, drops \cite{ho75,griggs07,sarrazin08}, lipid vesicles \cite{bruinsma96,vitkova04,coupier08,mcwhirter09,mcwhirter11}, capsules \cite{risso06,queguiner97,lefebvre08,lefebvre07,kuriakose11,hu11} or polymers \cite{reddig11} is often considered. Underlying motivations include a better understanding of blood flow, or the possibility to manipulate these objects in microfluidic devices for lab on chip applications. The most commonly reported stationary shapes are axisymmetric bullet-like and parachute-like shapes, the latter being characterized by a concave rear part. The shape of this rear part is very sensitive to the mechanical environment: therefore, observing it is a (cheap) rheology experiment in itself, as exemplified in Refs. \cite{queguiner97,lefebvre08}, where possible membrane constitutive laws for capsules are discussed. Similarly, in Ref. \cite{lefebvre07}, the onset of curvature inversion is shown to be strongly dependent on the capsule's pre-stress.

Alternatively, this shape will give indications on the hydrodynamic stresses on the object. It is generally observed that  increasing the flow velocity, or the confinement, leads first to an increase in the fore-aft asymmetry, then to the apparition of a negative curvature region at the rear, and eventually, at least for membrane-less objects, to break-up. Hydrodynamic interactions between neighboring objects are also strongly correlated to shape-dependent modifications of the local flow. In Ref. \cite{mcwhirter11}, small clusters of RBCs are simulated and two well separated states (compact or loose clusters) are explicitly associated with two different cell shapes (shallow or deep parachute). This hydrodynamic cell aggregation is mediated by a loop of fluid recirculation between cells, also called bolus \cite{gaehtgens80,pozrikidis05,mcwhirter09}, whose apparition or disappearance  is intimately correlated with shape changes \cite{mcwhirter11}. Thus, understanding the conditions for the apparition and stability of shapes and the resulting flow patterns around them provides a valuable entry point to build up hydrodynamic aggregation rules in a suspension. At larger scale, this self-organization of the suspension leads to specific rheological properties like the shear-thinning of blood or the F\aa rh\ae us-Lindquist effect \cite{fung93}. Even in the case of an isolated cell in a dilute suspension, shape variations will modify viscous dissipation, and therefore, the effective viscosity of  the suspension. Again, specific changes in this viscosity can be associated with concavity changes at the rear and loss of  membrane tension \cite{bruinsma96}.

In this paper, we identify shape changes  with  flow variations in the case of fluid vesicles, which are also simplified models for RBCs. We shall see  that the general and therefore intuitive sketch of concavity increase with hydrodynamic  stress must be reconsidered in the case of vesicles, which differ from drops or capsules by their inextensible fluid  membranes.  We study explicitly the effect of three flow characteristics: its velocity, the confinement, and its axial asymmetry.

\begin{figure}[b!]
  \includegraphics[width=\columnwidth]{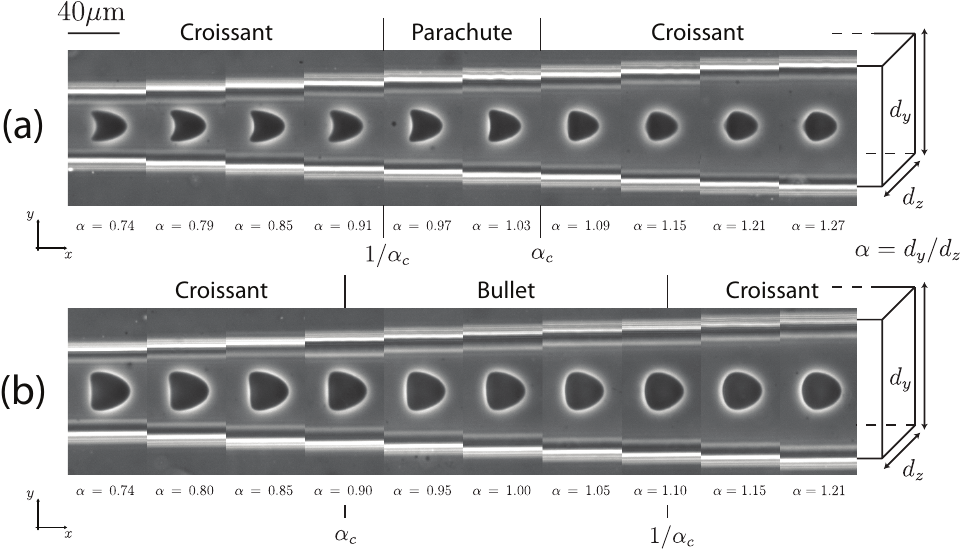}
\caption{Vesicle cross-sections in a channel of increasing aspect ratio. (a): $\nu=0.913$, $\hat{R}=0.35$,  $C_a=59$. (b): $\nu=0.973$, $\hat{R}=0.38$,  $C_a=81$. $\hat{R}$ and $C_a$ are given for the square section.\label{fig:typique}}
\end{figure}

\begin{figure}[t!]
\includegraphics[width=0.9\columnwidth]{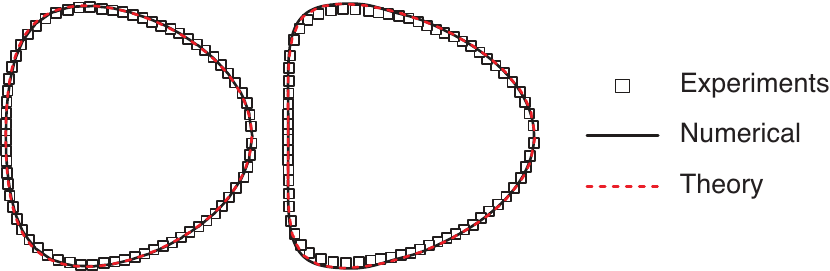}
\caption{\label{fig:compar}Cross-sections for two vesicles in symmetric flow (exp.: $\hat{R}=0.38$, sim. and th.: $\hat{R}=0$). Left: $\nu=0.979, C_a=4191$; right: $\nu=0.946, C_a=3776$. No fitting parameters.}
 \end{figure}

The problem is considered through experiments, numerical simulations, and theoretical calculations.
In the experiments, we use a PDMS microfluidic device. Vesicles are prepared following the standard electroformation method, which produces vesicles of various size and deflation. They are made of a DOPC lipid bilayer enclosing an inner solution of sugar in water.  Vesicles are then diluted in another sugar solution and imaged by phase contrast microscopy. The viscosity ratio between both solutions is close to unity. Vesicles flow  along the $x$ axis in a straight channel of constant thickness $d_z$  ($z$ direction) and varying width $d_y$ ($y$ direction). Their cross-section  in the $xy$ plane is observed (Fig. \ref{fig:typique}). Gravity is in the $x$ direction, so that quick centering is achieved \cite{coupier08}. Each section of given width is long enough for stationary shapes to be  reached. The analytical calculation is based on the decomposition of the  shape in spherical harmonics. For simplicity, we neglect the influence of channel walls and use for  velocity profile $v(y,z)=V\left[1-\left(2y/d_y\right)^2-\left(2z/d_z\right)^2\right]$, which turns out to be a valid approximation for low confinements. As in Ref. \cite{barthesbiesel80}, solving the Stokes equations together with the boundary conditions at the membrane and at infinity and using the condition of stationary shape, we find expressions for the amplitude of each considered harmonic as a function of $\nu,$ $C_a,$ and $\alpha$. This method is much more efficient than the traditional one, which is based on derivation and numerical solution of the shape evolution equations \cite{farutin11}. We can now take 18 harmonics for axisymmetric case and 12 for elliptic cross-sections. The accuracy of the results was verified by 3D numerical simulations using boundary integral method \cite{biben11}. Finding theoretical and numerical approaches that fit the experimental results up to the shape details is still a challenging issue, as illustrated by the acute debate around phase diagrams of vesicles under shear flow \cite{deschamps09,biben11}. Here, we find excellent agreements, as exemplified by the superimposition of shapes in Fig. \ref{fig:compar}.

Fluid vesicles have constant volume and  surface area. Their deformability is directly linked to their initial (and constant) deflation, given by the reduced volume $\nu=\mathcal{V}/\big[4\pi(\mathcal{S}/4\pi)^{3/2}/3\big]$, where $\mathcal{V}$ and $\mathcal{S}$ are the vesicle volume and membrane area. In the experiments, both are calculated thanks to axial symmetry of the shape in the square cross-section channel, on which we comment later on. The typical size of a vesicle is given by its effective radius $R=\sqrt{\mathcal{S}/4\pi}$. The geometry of the problem is characterized by two dimensionless numbers: the channel aspect ratio $\alpha=d_y/d_z$ and the confinement $\hat{R}=2 R/\sqrt{d_y d_z}$.  The capillary number, that compares viscous stress to membrane elasticity, is defined by $C_a=V/(d_y d_z) R^4 \eta /\kappa$, where $V$ is the unperturbed fluid maximum velocity, $\eta$ the fluid viscosity and $\kappa$ the membrane bending modulus.

Overall, the problem is described by the four dimensionless parameters $(\nu,\alpha,\hat{R},C_a)$. The explored ranges are summarized in Table \ref{tab:param}. 

\paragraph{Preliminary observation.}--- When vesicles flow from narrow to wide sections (the thickness $d_z$ being constant), their in-plane section becomes less asymmetric between the front and the rear, and the negative curvature region at the rear, if any, disappears (Fig. \ref{fig:typique}).  The shapes are symmetric about the $Ox$ axis \cite{note}.  A given vesicle flowing several times back and forth in the square channel always shows the same 2D shape, which indicates that it adopts, at least, the square symmetry. Thanks to digital holographic microscopy, we have recently seen that the vesicle's transverse cross-sections are elliptical  \cite{minetti08}. We then reach the a priori nontrivial conclusion that the vesicle shape obeys axial symmetry in a square geometry. We observe two types of axisymmetric shapes, namely parachute and bullet.
We first investigate their existence domains in symmetric channel ($\alpha=1$) and  then explore the influence of  asymmetry for low constraints $(\hat{R} \le 0.5,C_a\lesssim 500)$.

\begin{table}
\caption{\label{tab:param}Parameter ranges. For the experiments, $\eta = 10^{-3}$Pa.s, $\kappa= 10^{-19}$J and $V\in [10 ;8300]\mu$m.s$^{-1}$.} 
\begin{tabular}{|c|c|c|c|c|}
\hline
&$\nu$ & $\alpha$&$\hat{R}$&$C_a$\\
\hline
Experiments & $0.910 - 1$&$0.49 - 1.74$&$0.12 - 1.27$&$3 - 7\times 10^4$\\
Theory & $0.9 - 1$&$0.5 - 2$&0&$10 - 6\times 10^4$\\
\hline
\end{tabular}
\end{table}

\begin{figure*}[t!]
  \includegraphics[width=2\columnwidth]{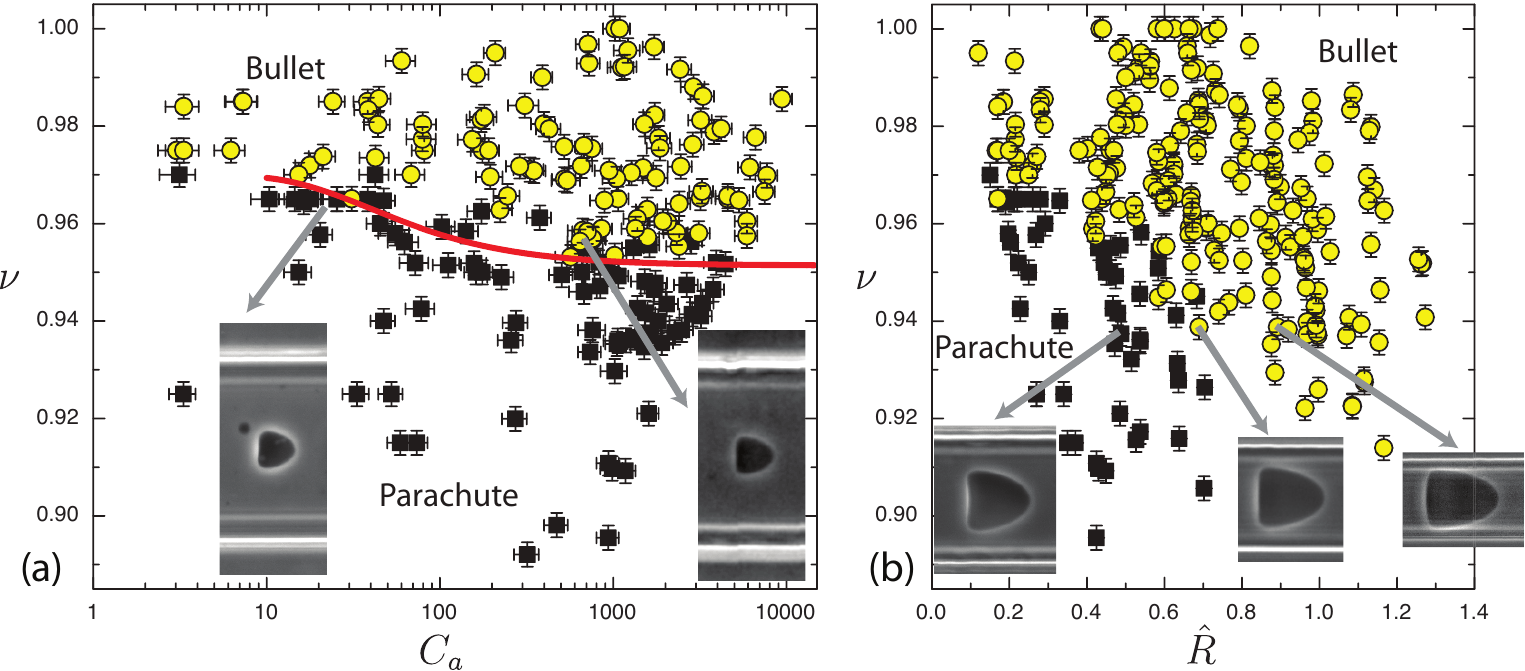}
\caption{(color online) Shapes for $\alpha=1$. Dots refer to experimental data and full line to theory and simulations. (a): bullet-parachute phase diagram in the $(\nu,C_a)$ plane for not too confined vesicles ($0.12\le \hat{R}\le 0.5$). Pictures in inset ($\nu=0.964, \hat{R}=0.24, C_a=20$ and $\nu=0.959, \hat{R}=0.25, C_a=667$) illustrate the concavity change as $C_a$ increases. (b): bullet-parachute phase diagram in the $(\nu,\hat{R})$ plane. For $\hat{R}\ge 0.4$, $C_a$ goes from 15 to 70000, while for $\hat{R}\le 0.4$, values are limited to the $C_a<100$ range due to shape dependency on $C_a$ in the unbounded case. Pictures in inset  ($\nu=0.938, \hat{R}=0.49, 0.69$ and $0.89$) illustrate the concavity change as $\hat{R}$ increases.\label{fig:parbullet}}
\end{figure*}

\paragraph{Effect of $C_a$ and $\hat{R}$ in symmetric channel.}---  For weakly confined vesicles ($\hat{R}\le0.5$), clear separation between the bullet and parachute domains is achieved in the $(\nu,C_a)$ space, although $\hat{R}$ varies by a factor 4 (Fig. \ref{fig:parbullet}(a)). A completely unexpected observation is the crossover from parachute to bullet by increasing $C_a$ for vesicles with a reduced volume between 0.95 and 0.97. The same trend is theoretically observed for unbounded flow. The full 3D simulations support these results. We observe, however, a slight shift in $C_a$ between experiments and theory. This could be attributed to uncertainties on the bending modulus (not measured) and to effects of thermal fluctuations  and confinement, which were neglected  in the theory. We have checked that taking into account a possible spontaneous curvature of the membrane has only a minor effect on the results.

While, for weak confinement ($\hat{R}\le 0.5$), the parachute to bullet crossover depends on the capillary number but not on the confinement, the situation is completely opposite for more confined vesicles. As shown in Fig. \ref{fig:parbullet}(b), for $\hat{R}\ge 0.5$, well separated domains are found in the $(\nu,\hat{R})$ plane, despite the strong variations of $C_a$  (more than three decades). This indicates that when confinement is  large enough, its effect  is dominant. Contrary to the intuition, the bullet shape is favored upon an increase of confinement.

\paragraph{Effect of asymmetry.}--- We observed that, for a vesicle of given reduced volume $\nu$, the 2D cross-section, and in particular the concavity, is independent  of $\hat{R}$ and $C_a$ for low constraint  ($\hat{R}\le 0.5$, $C_a\lesssim500$), and depends only on asymmetry $\alpha$ \cite{suppl}. Due to this $(\hat{R},C_a)$ independence, the full 3D shape of a vesicle under flow of aspect ratio $\alpha$ is deduced from its two in-plane shapes in the sections of aspect ratios $\alpha$ and $1/\alpha$, where $\hat{R}$ and $C_a$ are different. For instance, in Fig. \ref{fig:typique}(a), the shapes in sections $\alpha=0.79$ and $\alpha=1.27\simeq1/0.79$ can be seen as the top and side views of the same vesicle in a rectangular channel  of aspect ratio 0.79.  We define $\alpha_c$ as the aspect ratio of the channel where the crossover from concave to convex 2D shape occurs. If $\alpha_c$ is larger  than 1, as in Fig. \ref{fig:typique}(a), then for $1\le \alpha < \alpha_c$, the in-plane cross-section of the vesicle is concave, and its  orthogonal cross-section, for which we consider $1/\alpha< \alpha_c$, is also concave.
\begin{figure}[t!]
  \includegraphics[width=\columnwidth]{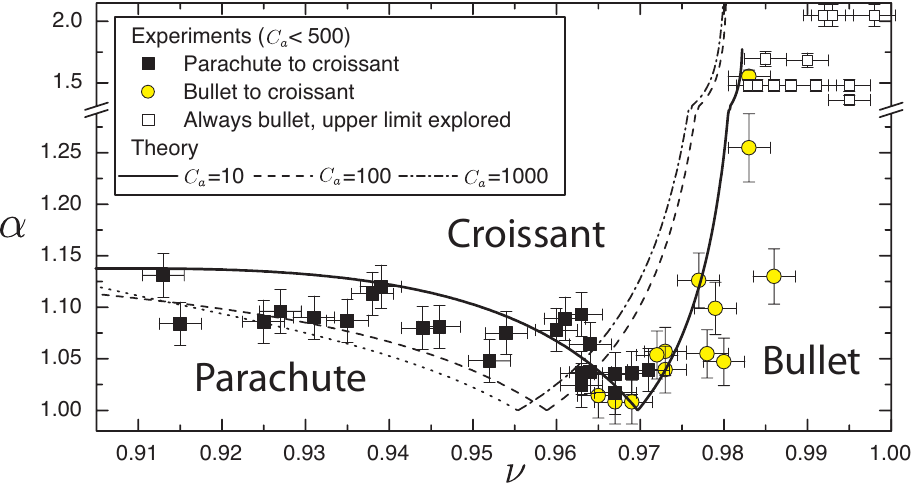}
\caption{Shape diagram in asymmetric channel at low constraint. Exp.: each point corresponds to a vesicle with $\hat{R}\le0.5$  and $C_a\lesssim 500$ when $\alpha=1$. Th.: $\hat{R}=0$, $C_a\le1000$.\label{fig:diagram}}
\end{figure}
The 3D shape  is therefore a parachute. If $\alpha\ge \alpha_c$, the in-plane shape is convex, while the orthogonal shape is concave,  since $1/\alpha\le\alpha_c$. Such a "croissant" shape has been seen for drops between two infinite planes \cite{griggs07}. If $\alpha_c$ is lower  than 1, as in Fig. \ref{fig:typique}(b), the same reasoning shows that the 3D shape is a bullet for $1\le \alpha \le 1/\alpha_c$ and a croissant for $\alpha > 1/\alpha_c$. Some weakly deflated vesicles show no concave shape in the explored $\alpha$ range, and are therefore always bullets. Combining all these data, we deduce the general phase diagram in the $(\nu,\alpha)$ space (Fig. \ref{fig:diagram}, restricted to $\alpha\ge 1$ for symmetry reasons), which is in good agreement with the theory for $\hat{R}=0$ and low $C_a$. For $\alpha=1$, one gets a parachute or a bullet shape; increasing the asymmetry of the channel leads to the croissant shape, which has a concave part in the plane of higher stress.

\begin{figure}[t!]
\includegraphics[width=0.9\columnwidth]{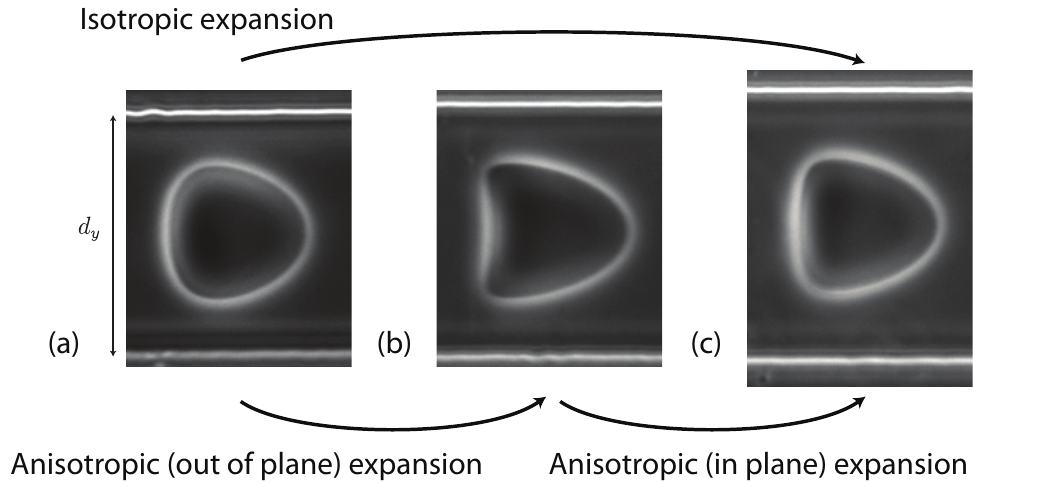}
\caption{\label{fig:zdirection}Cross-sections for the same vesicle ($\nu=0.985$) in three channels. (a): $d_y=d_z=83\mu$m; (b): $d_y=83\mu$m, $d_z=92\mu$m; (c): $d_y=d_z=92\mu$m. (a) and (b) give indications on what would be seen in the $xz$ plane of Fig. \ref{fig:typique}(b).}
 \end{figure}

\paragraph{Discussion and conclusion.} ---  We reported the phase diagram of vesicle shapes in Poiseuille flow in the relevant parameter space. Our study reveals the necessity to distinguish between isotropic and anisotropic  variations of the cross-section of the channel. While the evolution shown in Fig. \ref{fig:typique} is consistent with expectations (increasing confinement leads to more curved shapes), increasing isotropically the confinement leads actually to less curved shapes, as shown by shapes in inset of Fig. \ref{fig:parbullet}(b). The same conclusion is reached if the flow strength is increased at a given confinement. Theoretical and numerical results support these conclusions, for symmetric but also for asymmetric flows: as shown in Fig. \ref{fig:diagram}, the less distorted bullet and croissant shapes are favored against the parachute shape upon increasing flow strength.
Fig. \ref{fig:zdirection} summarizes our main finding; isotropic expansion (a-c) leads unexpectedly to more shape distortion while in-plane-only expansion (b-c) leads to a less curved shape. In that case,  the asymmetry between the two transverse directions allows to relax the constraints of perimeter and surface conservation at the level of a given longitudinal cross-section, whose evolution is therefore similar to the one of a drop or a capsule. Finally, the predominant role played by the aspect ratio of the channel is illustrated by the concavity increase through out-of-plane-only expansion (a-b), which then appears to be equivalent to an in-plane contraction. The latter case also indicates that a sole 2D view in experiments might lead to biased considerations when discussing the existence of parachutes, as in Ref.  \cite{noguchi10_2}.

\begin{figure}[t!]
\includegraphics[width=0.9\columnwidth]{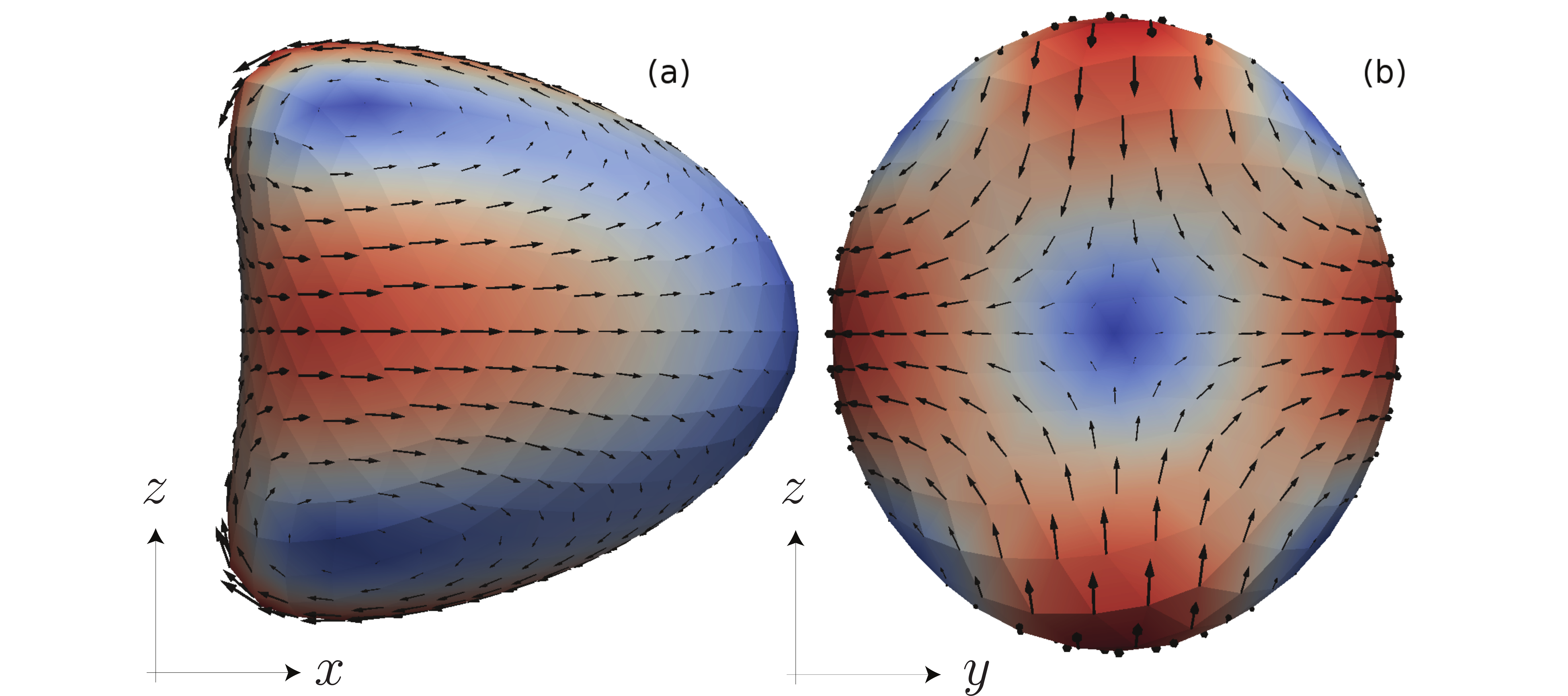}
\caption{\label{fig:tanktreading}(color online) Velocity field in the vesicle co-moving frame. Side (a)  and 
rear (b)  view. Simulations, $\nu=0.95,$ $C_a=100,$ $\alpha=1.2$ (croissant shape). The norm of velocity
 is color-coded. Maximum velocity is about 5 times lower than vesicle velocity. See also \cite{suppl}.}
\end{figure}

In asymmetric channels, vesicle shapes exhibit another unexpected feature: as seen in Fig. \ref{fig:typique} and in \cite{suppl}, their in-plane width does not increase with $d_y$ but is roughly constant. Therefore, contrary to drops \cite{sarrazin08}, vesicles do not adopt the aspect ratio of the channel but  occupy a space in the $yz$ plane that is delimited by a circle. 
The fact that the fluid then has to go through a small gap in the narrowest width of the channel, may lead to the intuitive but wrong conclusion that the vesicle should decrease its extent in that direction (and follow more or less the channel aspect ratio). In order to get an insight  on this peculiar phenomenon, we investigated the patterns of membrane surface flow through simulations  (Fig. \ref{fig:tanktreading}). As soon as the channel is asymmetric, 4 vortices appear on the surface with a backward flow in the direction corresponding to the narrowest gap, which decreases the mean shear in the gap and the viscous stress on the membrane. The opposite takes place in the other direction, leading to a more homogeneous stress distribution.
When drops adopt  the channel cross section, all material points on their surface are advected backward \cite{suppl}. The corresponding flow line possess two in-plane stagnation points at
the front and at the back, on the main axis. This apparent singularity is resolved for drops through recirculation
inside the drop, a fact that is precluded for a membrane  whose material points must stay on the surface. This difference has already been underlined for drops and vesicles bounded to a substrate and submitted to shear flow  \cite{vezy07}.  The 4-vortex pattern is thus the simplest acceptable flow pattern on the vesicle surface, if we consider that zero flow situation is unlikely to happen for this fluid membrane under asymmetric constraint. This pattern can be compared to the one obtained recently for vesicle sedimentation \cite{boedec12}. Despite the axial symmetry of the problem, it is shown that  a croissant shape with four vortices on its surface is a possible stationary shape. However, the stability of this solution is not discussed.


Flow asymmetry, which can be due to the channel geometry or to the presence of neighboring cells, will have therefore two consequences: from the rheological point of view, surface vortices of non negligible velocity (see Fig. \ref{fig:tanktreading}) imply important motions of the fluid inside the vesicle (or the RBC), that will contribute to the net dissipation. These vortices, that would not be caught for instance by 2D simulations, will also affect accordingly the flow field around the membrane, which will modify in return the recirculating loop between cells. Whether this mechanism would be a stabilizing or destabilizing factor in cell clusters, that often exhibit symmetry breaking \cite{mcwhirter09},  remains to be discussed.

\begin{acknowledgments}
We acknowledge financial support from CNES and the ANR "MOSICOB project". This work was also supported by the SSTC/ESA-PRODEX (Services Scientifiques Techniques et Culturels/European Space Agency - Programmes de D\'eveloppement d'exp\'eriences) contract 90171.

\end{acknowledgments}

\newpage

$\,$

\newpage

\begin{center}
\textbf{Supplemental material for "Shape diagram of vesicles in Poiseuille flow"
}\end{center}

In this supplemental material, details of shape changes with channel aspect ratio $\alpha$ in the experiments are given. We also show more velocity fields on the membranes of vesicles  and compare them with the case of droplets.

\section{Details of shape variations}

\begin{figure}[h]
\resizebox{\columnwidth}{!}{\includegraphics{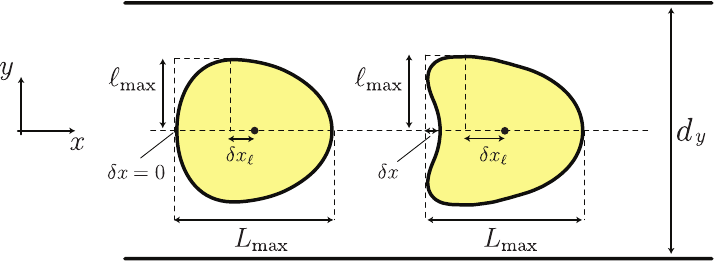}}
\caption{(color online) The different geometrical parameters for the in-plane cross-section of a vesicle (convex or concave).} \label{fig:formes}
\end{figure}

Due to the symmetry with respect to the $x$ axis, we mainly characterize the in-plane shapes by two parameters (see  Figure~\ref{fig:formes}): the concavity at the rear  $c=\delta x/L_{\max}$ and the fore-aft asymmetry $a_s=2 \delta x_{\ell}/L_{\max}$.

\begin{figure*}[t!]
\resizebox{2\columnwidth}{!}{\includegraphics{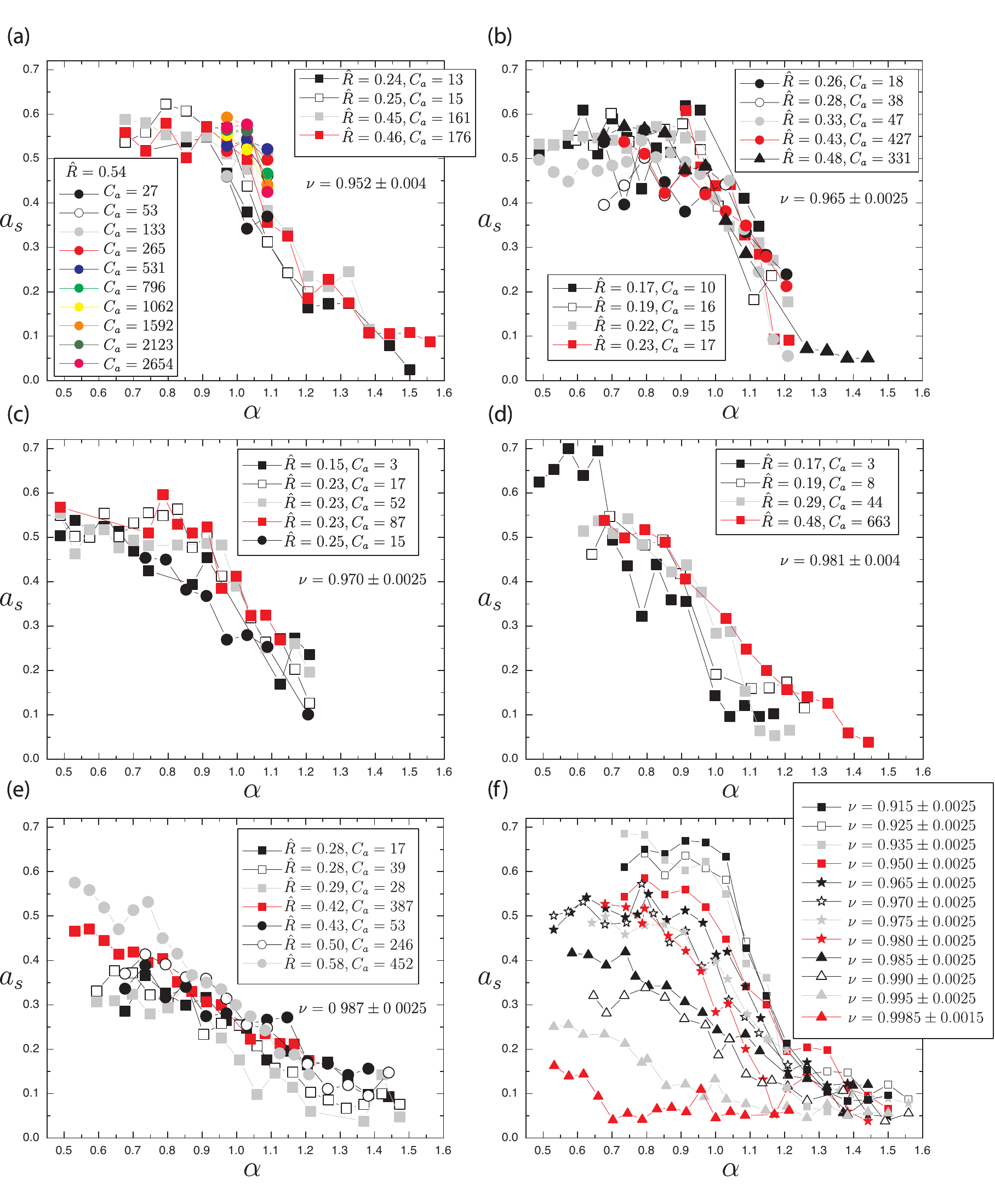}}
\caption{(color online) (a-e): in-plane asymmetry  $a_s$ as a function of aspect ratio $\alpha$ for several vesicles in the same tiny interval of reduced volume. $\hat{R}$ and $C_a$ are given for the square cross-section. (a): $\nu=0.952\pm0.004$ ; (b): $\nu=0.965\pm0.0025$ ; (c): $\nu=0.970\pm0.0025$ ; (d):  $\nu=0.981\pm0.004$ ; (e): $\nu=0.987\pm0.0025$. (f): in-plane asymmetry  $a_s$ as a function of aspect ratio $\alpha$ for different reduced volumes (averaged over several representative vesicles with $\hat{R}\le0.5$ and $C_a<500$, from 1 to 10 vesicles for each curve).$ \, $ $ \, $ $ \, $ $ \, $ $ \, $ $ \, $ $ \, $ $ \, $ $ \, $ $ \, $ $ \, $ $ \, $ $ \, $ $ \, $ $ \, $ $ \, $ $ \, $ $ \, $ $ \, $ $ \, $ $ \, $ $ \, $ $ \, $ $ \, $ $ \, $ $ \, $ $ \, $ $ \, $ $ \, $ $ \, $ $ \, $ $ \, $ $ \, $ $ \, $ $ \, $ $ \, $ $ \, $ $ \, $ $ \, $ $ \, $ $ \, $ $ \, $ $ \, $ $ \, $ $ \, $ $ \, $ $ \, $ $ \, $ $ \, $ $ \, $ $ \, $ $ \, $ $ \, $ $ \, $ $ \, $ $ \, $ $ \, $ $ \, $ $ \, $ $ \, $ $ \, $ $ \, $ $ \, $ $ \, $ $ \, $ $ \, $ $ \, $ $ \, $ $ \, $ $ \, $ $ \, $ $ \, $ $ \, $ $ \, $ $ \, $ $ \, $ $ \, $ $ \, $ $ \, $ $ \, $ $ \, $ $ \, $ $ \, $ $ \, $ $ \, $ $ \, $ $ \, $ $ \, $ $ \, $ $ \, $ $ \, $ $ \, $ $ \, $ $ \, $ $ \, $ $ \, $ $ \, $ $ \, $ $ \, $ $ \, $ $ \, $ $ \, $ $ \, $ $ \, $ $ \, $ $ \, $ $ \, $ $ \, $ $ \, $ $ \, $ $ \, $ $ \, $ $ \, $ $ \, $ $ \, $ $ \, $ $ \, $ $ \, $ $ \, $ $ \, $ $ \, $ $ \, $ $ \, $ $ \, $ $ \, $ $ \, $ $ \, $ $ \, $ $ \, $ $ \, $ $ \, $ $ \, $ $ \, $ $ \, $ $ \, $ $ \, $ $ \, $ $ \, $ $ \, $ $ \, $ $ \, $ $ \, $  } \label{fig:asym}
\end{figure*}

\begin{figure*}[t!]
\resizebox{2\columnwidth}{!}{\includegraphics{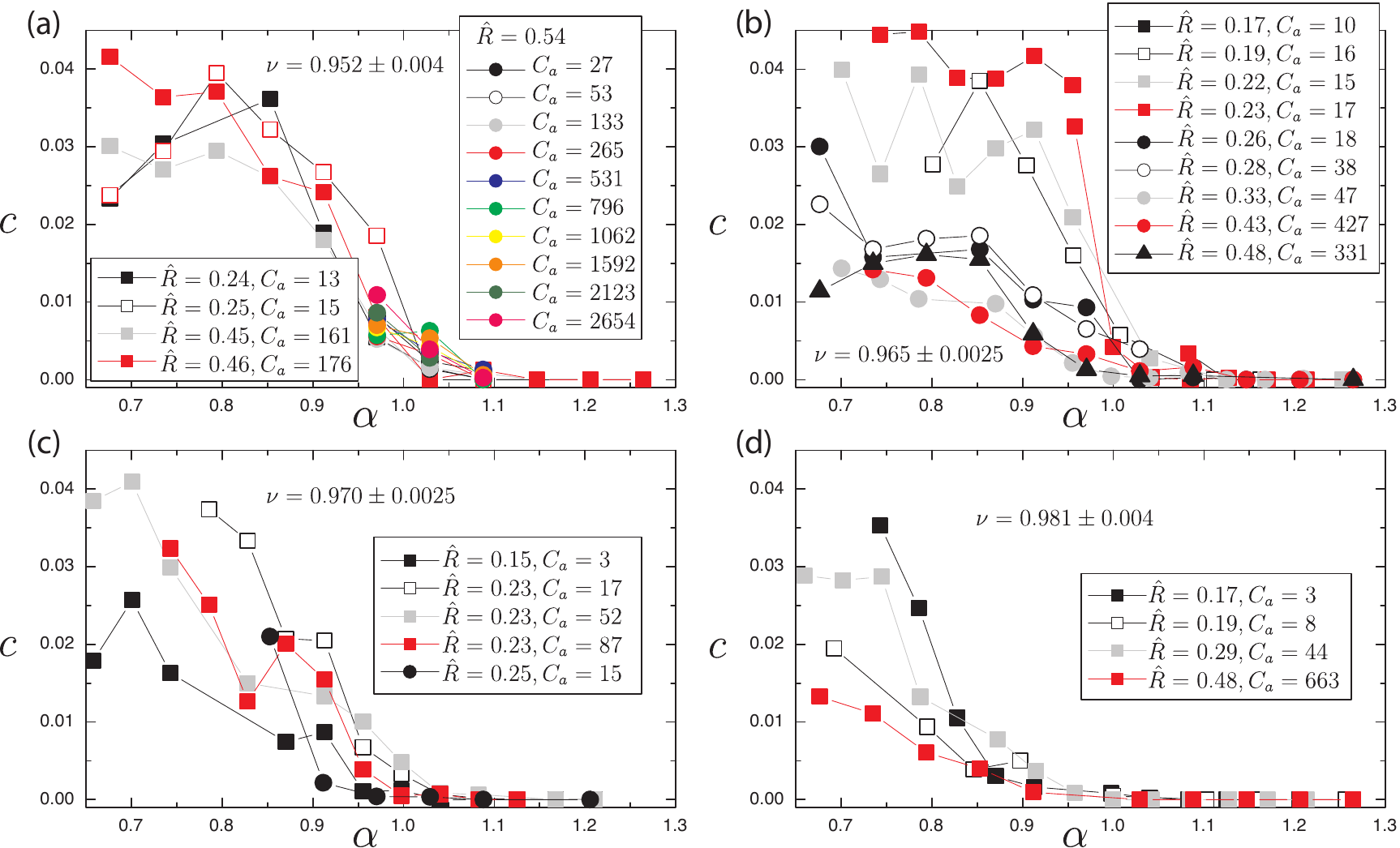}}
\caption{(color online) In-plane concavity $c$ as a function of aspect ratio $\alpha$ for several vesicles in the same tiny interval of reduced volume. $\hat{R}$ and $C_a$ are given for the square cross-section. (a): $\nu=0.952\pm0.004$ ; (b): $\nu=0.965\pm0.0025$ ; (c): $\nu=0.970\pm0.0025$ ; (d):  $\nu=0.981\pm0.004$. The decrease of critical $\alpha_c$ with increasing $\nu$ is noticable.} \label{fig:conc}
\end{figure*}

In the experiments, vesicles of varying reduced volume $\nu$ and radius $R$ were observed in channels of constant thickness $d_z$ and varying width $d_y$ under a flow of maximum velocity $V$ (see pictures of Figures 1 in the main paper). For a selection of tiny intervals of reduced volumes, we plot in Figures \ref{fig:asym}(a-e) and \ref{fig:conc}(a-d) the variations of $a_s$ and $c$ with $\alpha$ for vesicles under different confinement $\hat{R}$ and capillary number $C_a$ that is varied either by changing $V$ or because of its $\hat{R}$ dependency. In the data shown in Figures \ref{fig:asym}(a-e) and \ref{fig:conc}(a-d), $\hat{R}$  varies by a factor higher than $3$ and $C_a$ by a factor 100. In particular, in Figures  \ref{fig:asym}(a) and \ref{fig:conc}(a), the same vesicle with $\hat{R}=0.54$ in the square cross-section channel is observed around $\alpha=1$ for velocities $V$ varying by a factor 100, so that $C_a$ goes from 27 to 2654 in the square cross-section channel. In Figures  \ref{fig:asym}(c) and \ref{fig:conc}(c), the same vesicle with $\hat{R}=0.23$ is observed on a wider  interval of $\alpha$ for velocities $V$ varying by a factor 5, so that $C_a$ goes from 17 to 87.

For a given $\nu$, the curves $a_s(\alpha)$  collapse  reasonably well on a single one. In particular, $a_s$ varies by a factor 5 or 6 while $\alpha$ varies typically from  1 to 1.5 while the residual variations around the master curve are much smaller even though $\hat{R}$ varies by a factor 3. This indicates that the asymmetry is controlled by the aspect ratio but not by the confinement (within the limit $\hat{R}\lesssim 0.5$). In the range $C_a\lesssim 500$, it seems also that the $C_a$-dependency is weak. These facts suggest that the mean stress on the membrane (which is linked to the mean confinement or to fluid velocity) is not important, and that the stress distribution around the vesicle is the key parameter.

Finally, figure \ref{fig:asym}(f) shows the variations of asymmetry $a_s$ with $\alpha$ for different reduced volumes. Asymmetry is a decreasing function of $\alpha$: in-plane asymmetry is higher in the narrower longitudinal section of the channel. From the plateau observed at low $\alpha$, there seems to be a  maximal possible asymmetry, which increases with deflation of vesicle, as one would expect.

\begin{figure*}[h!]
\begin{center}
  \includegraphics[width=1.8\columnwidth]{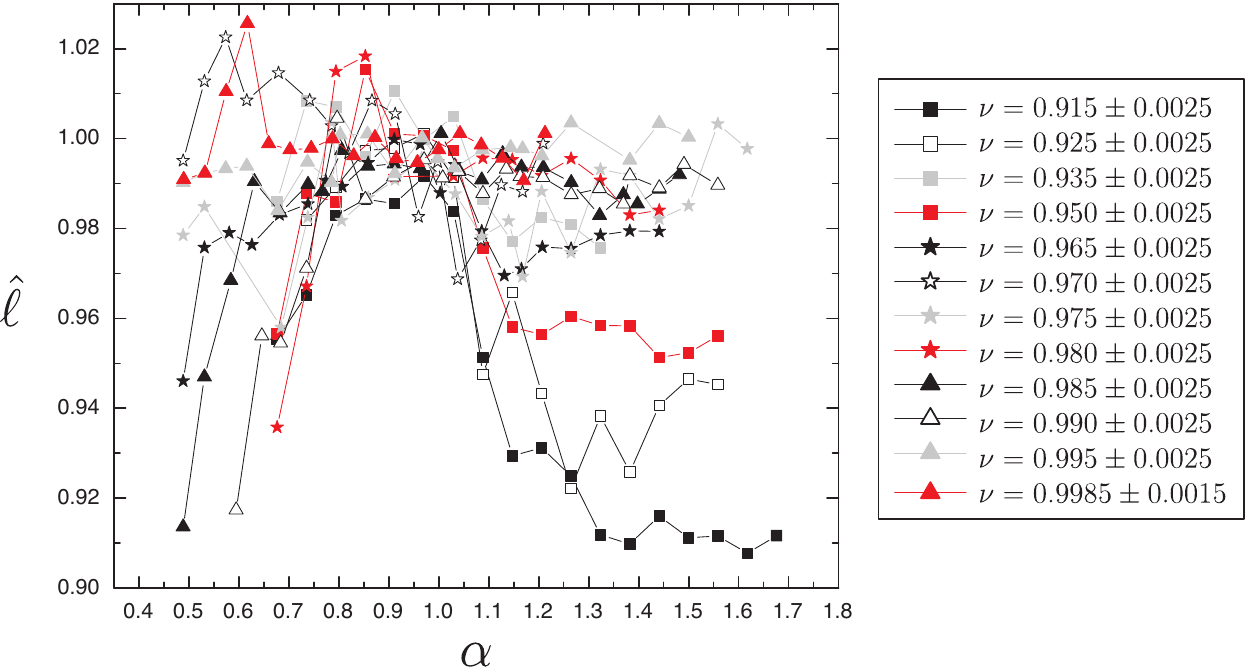}

\caption{Maximum width $\hat{\ell}=\ell_{\max}/R$ as a function of $\alpha$. The initial increase can probably be associated  with the proximity of walls (in plane confinement $R/d_y$ larger than 0.6). Note that the typical error bars on these data are around $2\%$, which is of the same order as the variations observed (but the slight decrease for high $\alpha$ for deflated vesicles)}\label{fig:ellmax}
\end{center}
\end{figure*}

 \begin{figure*}[t!]
\begin{center}
  \includegraphics[width=2\columnwidth]{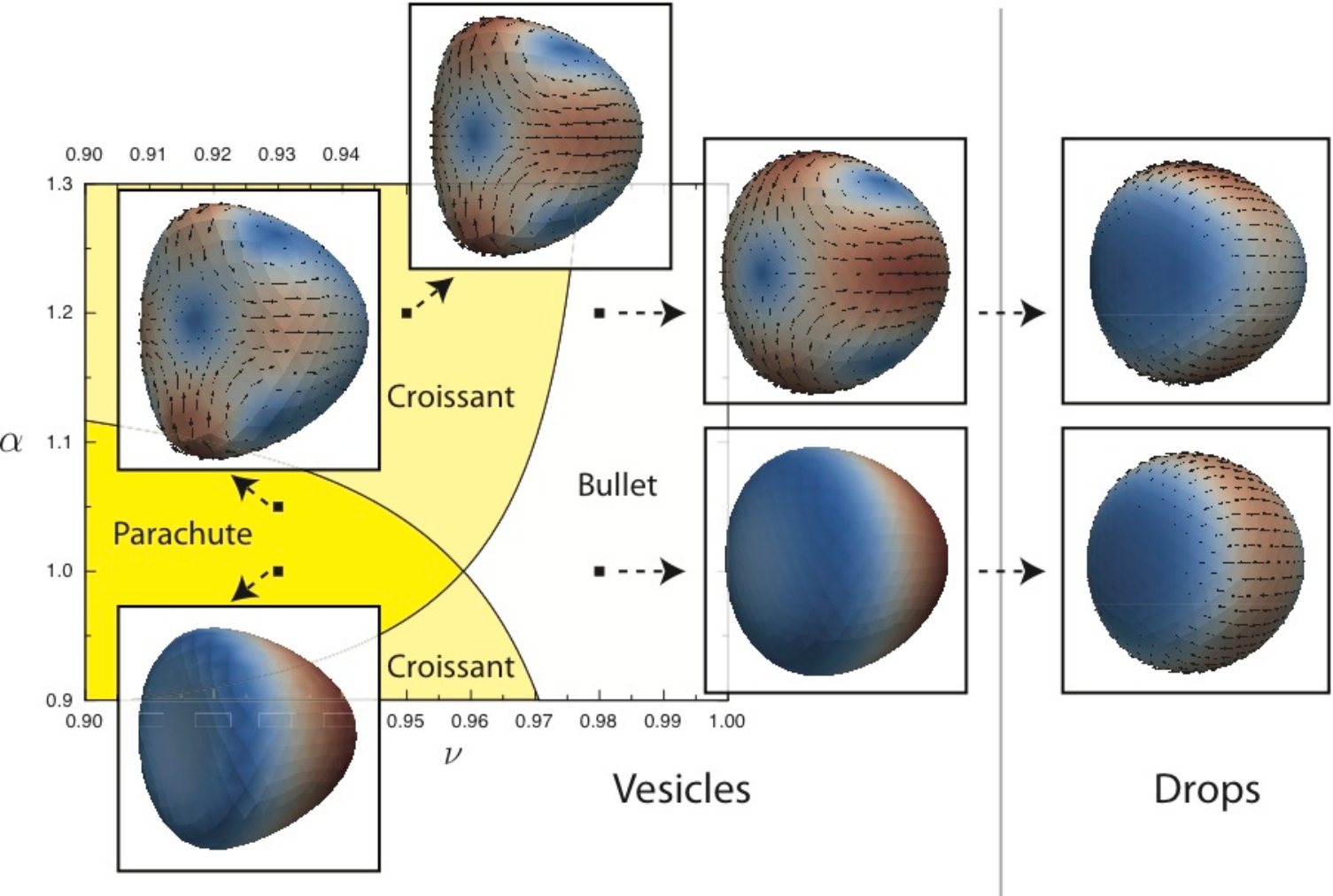}

\caption{3D theoretical shapes for vesicles and drops with fluid velocities at the surface. Color code shows the norm of the velocity, except for vesicles at $\alpha=1$ where there is no membrane motion and color code shows membrane tension. Vesicles : $C_a=100$. Drops : $C_a=0.2$.}\label{fig:shapes}
\end{center}
\end{figure*}

Like asymmetry, concavity is controlled by the aspect ratio of the channel since the curves $c(\alpha)$  fall down to 0 at the same critical $\alpha_c$ (concave $\to$ convex crossover). For $\alpha\le\alpha_c$, no complete collapse is observed, but this can be linked to the fact that tension in the concave rear part is weak and the membrane is therefore submitted to higher thermal fluctuations and might be more sensitive to variations in the fluid velocity $V$.\\

A vesicle shape is also characterized by its reduced width $\hat{\ell}=\ell_{\max}/\hat{R}$ (see Figure \ref{fig:formes}). As shown in Figure \ref{fig:ellmax}, $\hat{\ell}$ is more or less constant whatever $\alpha$. Surprisingly, deflated enough ($\nu\lesssim0.95$) vesicles are even narrower when the aspect ratio increases, so when the in-plane confinement decreases. Thus, in general, increasing the confinement does not lead to a in-plane shrinkage of the vesicle, so that the membrane is submitted to a lower flow stress, but only displaces the location $a_s$ of the maximum thickness to the rear .

These results imply that, at the rear, the aspect ratio of the transverse cross-section of the vesicle is reversed compared to the channel aspect ratio. It is equal to 1 somewhere in the middle and, at the front, follows the same aspect ratio as the channel.
These results lead to some comments on a previous study: in Ref. \cite{minetti08} we observed vesicles flowing in a channel of square cross-section. Vesicles were observed by digital holographic microscopy, which allowed us to measure the optical thickness $e(y,x)=2\Delta n(y,x) z(y,x)$, where $z$ is the membrane position and $\Delta n$ the refractive index difference between the inner and the outer fluid. Because $\alpha$ was equal to 1, we made the assumption of axisymmetry of the vesicle and from this information on $z$ we deduced values for the refractive index difference along the vesicle $\Delta n (x)$. 

Strong gradients were observed, that increased with vesicle velocity and could lead to an increase by a factor 2 of the refractive index difference between the rear and the front of the vesicle, leading to the conclusion of the existence of huge gradients of  sucrose concentration inside the vesicle. However, it was argued that, according to the permeability values found in the literature,  difference of osmotic pressure across the membrane should prevent such important concentration gradients, even though residual flow across the permeable membrane due to hydrodynamic pressure difference could effectively lead to some sugar advection inside the vesicle (whose membrane is not permeable to sugar). The observed gradients in $e$ were nevertheless attributed to concentration gradients because it was assumed that the vesicle shape should follow the channel's aspect ratio, even in case of small departure from  the square section case.

Indeed, in these experiments, poor channel quality led us to estimate the uncertainty on $\alpha$ to around 3\%, that is, $0.97\lesssim \alpha\lesssim 1.03$. We now know that it  can lead to important variations of the vesicle's aspect ratio along the $x$ axis, because, as seen in Figure \ref{fig:asym}(e), $a_s$ variations are particularly strong around $\alpha=1$, while $\hat{\ell}$ remains quasi constant. Part of the observed gradients in Ref. \cite{minetti08} could therefore be attributed to these aspect ratio variations.

\section{Fluid motion at the surface}

In Figure \ref{fig:shapes} we show additional 3D shapes for vesicles at $C_a=100$. Tank-treading velocities at the surface are shown. In the axisymmetric case, no movement is seen, but 4 vortices appear as soon as $\alpha\ne1$, for parachutes, bullets or croissants. For comparison, we calculated the shape and surface velocities for drops with no viscosity contrast and capillary number $C_a=0.2$. $C_a$ is defined by $C_a= \eta V R^2 / (\sigma d e)$, where $\sigma$ is the surface tension. Note that the choice of the value $C_a$ value is arbitrary (within the condition of low enough value to prevent break-up), as the very different nature of the interfaces makes any quantitative comparison between drops and vesicles tricky, even with similar capillary numbers.

The flow patterns on the drop surface are completely different, with a  front to rear movement even for $\alpha=1$, and no qualitative change in the asymmetric case, even if continuity of velocity field implies increased velocity in the plane of higher flow curvature.


\begin{thebibliography}{31}
\bibitem{note}
\bibinfo{note}{Asymmetric slippers are occasionally observed but are not stable. Stability conditions of slipper shapes for vesicles and RBCs are still under debate \cite{skalak69,kaoui09_2,farutin11,mcwhirter11} and we shall not enter into it much further.}
\end{thebibliography}
\end{document}